\documentclass{mem}
\usepackage{natbib}\usepackage{txfonts}\usepackage{balance}
\usepackage{graphicx}
\usepackage[a4paper,breaklinks,dvipdfm]{hyperref}
\idline{75}{282}
\begin{document}

\title{
Disentangling multiple stellar populations in globular clusters using the Str\"omgren system
}

   \subtitle{}

\author{
J.\, Alonso-Garc\'ia\inst{1} 
\and M.\, Catelan\inst{1,2}
\and P.\, Amigo\inst{3}
\and C.\, Cort\'es\inst{4}
\and C. A.\, Kuehn\inst{5}
\and F.\, Grundahl\inst{6}
\and G.\, L\'opez\inst{7}
\and R.\, Salinas\inst{8}
\and H. A.\, Smith\inst{9}
\and P. B.\, Stetson\inst{10}
\and A. V.\, Sweigart\inst{11}
\and A. A. R.\, Valcarce\inst{12}
\and M.\, Zoccali\inst{1}
          }

  \offprints{J. Alonso-Garc\'ia}

\institute{
Departamento de Astronom\'{i}a y Astrof\'{i}sica, 
Pontificia Universidad Cat\'{o}lica de Chile, 
Av. Vicu\~na Mackenna 4860, 782-0436 Macul,
Santiago, Chile
\and
The Milky Way Millennium Nucleus,
Santiago, Chile
\and
Universidad de Valpara\'iso,
Valpara\'iso, Chile
\and
Universidad Metropolitana de Ciencias de la Educaci\'on,
Santiago, Chile
\and
Sydney Institute for Astronomy, University of Sydney, 
Sydney, Australia
\and 
Aarhus University,
Aarhus, Denmark
\and
KU Leuven,
Leuven, Belgium
\and
Finnish Centre for Astronomy with ESO, University of Turku,
Turku, Finland
\and
Michigan State University, 48824,
East Lansing, MI
\and
Dominion Astrophysical Observatory,
Victoria, Canada
\and
NASA Goddard Space Flight Center,
Greenbelt, MD
\and
Universidade Federal do Rio Grande do Norte,
Natal, Brazil
 \\
\email{jalonso@astro.puc.cl}
}

\authorrunning{Alonso-Garc\'ia}

\titlerunning{Disentangling multiple stellar populations in GCs}

\abstract{ An increasing amount of spectroscopic and photometric
  evidence is showing that the stellar populations of globular
  clusters are not as simple as they have been considered for many
  years. The presence of at least two different populations of stars
  is being discovered in a growing number of globular clusters, both
  in our Galaxy and in others.  We have started a series of
  observations of Galactic globular clusters using the Str\"omgren
  photometric system in order to find the signatures of these multiple
  populations and establish their presence in a more complete sample
  of globular clusters in the Milky Way, and to study their radial
  distributions and extensions. We present here the first results of
  our survey.  
  \keywords{Stars: Hertzsprung-Russell and C-M diagrams
    -- Stars: abundances -- Stars: atmospheres -- Stars: Population II
    -- Galaxy: globular clusters -- Galaxy: abundances} }
\maketitle{}

\section{Introduction}

In the last years, the combination of big telescopes and
state-of-the-art spectrographs have allowed to obtain high-resolution
spectra of several tens, even a few hundreds, of stars in a
significant number of Galactic globular clusters (GCs; e.g.,
\citealt{ca09,jo12,co12,mu12}). These spectra have shown that
variations in light element composition among stars in the same
cluster seem to be the rule, and not the exception.  These variations
present themselves as anti-correlations of pairs of elements (e.g.,
Na-O, Mg-Al, C-N) and have allowed to separate individual cluster
stars in at least two different populations \citep{ca10}. These
multiple populations are believed to have formed in subsequent
star-formation episodes, where stars from the latest generations are
chemically enriched with respect to the first generation
(\citealt{va11}, and references therein). But the mechanism of
self-enrichment and its extension is still a matter of current
discussion and debate (\citealt{gr12}, and references therein).

Photometry can also help to explore the existence, proportion and
distribution of multi-populations in GCs. Deep photometric
observations of the Galactic GCs are usually less expensive in
telescope time than spectroscopic ones, which is an important
advantage. But observations in the most common optical filters only
show the effects of multiple populations when using very precise and
deep photometry \citep{pi07,mi08}. Observations using the Str\"omgren
filters, especially the ultraviolet band, have been suggested to be
more efficient in showing the effects of different stellar population,
due to sensitivity of their passbands to strong molecular bands such
as CN, NH, or CH (e.g., \citealt{gr02,yo08,sb11,ca11}).
%Observations in the ultraviolet have been suggested to be more efficient in showing the effects of different stellar populations in GCs \citep{sb11}, and have been proven so, both from the space \citep{mi11,pi12} and from the ground {ca11}.

\section{Our survey: observations and first results}

We have begun a survey to observe the effects of the presence of
multiple populations in Galactic GCs using Str\"omgren photometry. In
our survey we are using the 4.1m SOAR telescope, located at the Cerro
Pach\'on Observatory in Chile. We are performing our observations with
the SOI camera, in a configuration which provides us with a pixel
scale of $0.154''$ and a field of view (FOV) of
$5.25'\times5.25'$. Since the FOV is too small to cover the whole area
of the observed Galactic GCs, we are observing the GCs following a
mosaic pattern to optimize their spatial coverage. We are using four
Str\"omgren filters ($u$, $v$, $b$, and $y$), plus the Bessel $I$ for
a more complete wavelength coverage. We have been able to observe up
to now 30 Galactic GCs. We obtained the PSF photometry from the images
using an updated version of Dophot \citep{sc93,al12}. We are
calibrating the photometry using a set of GCs with previous
well-calibrated Str\"omgren photometry \citep{gr99}, and \citet{st00}
photometric standard stars in $I$. Also we have astrometrized our
observations by comparison with bright stars obtained in each field
from the Two Micron All Sky Survey (2MASS; \citealt{sk06}) catalog
available through the Infrared Processing and Analysis Center (IPAC)
website.

\begin{figure*}[t!]
\resizebox{\hsize}{!}{
\includegraphics[clip=true]{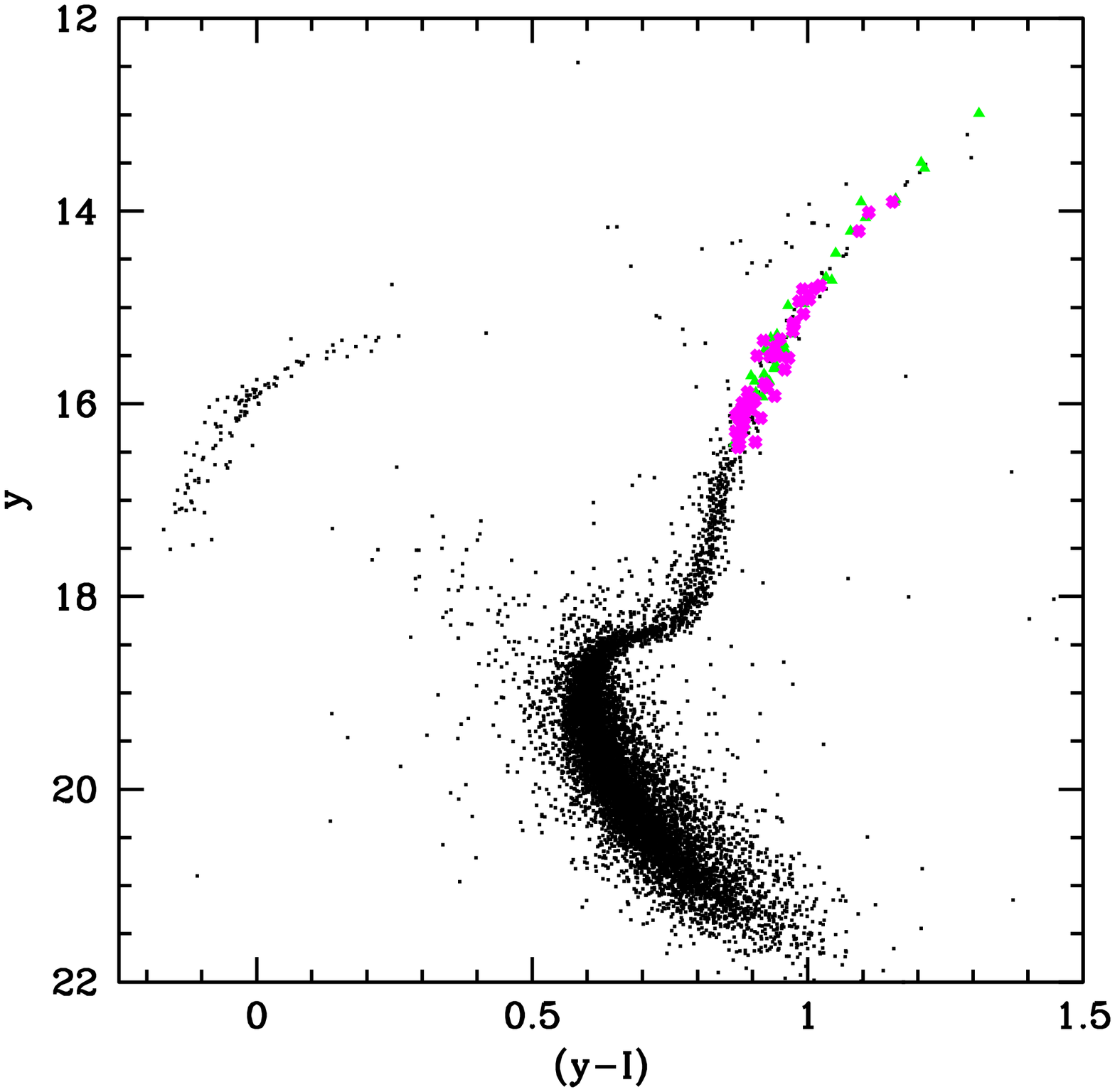}
\includegraphics[clip=true]{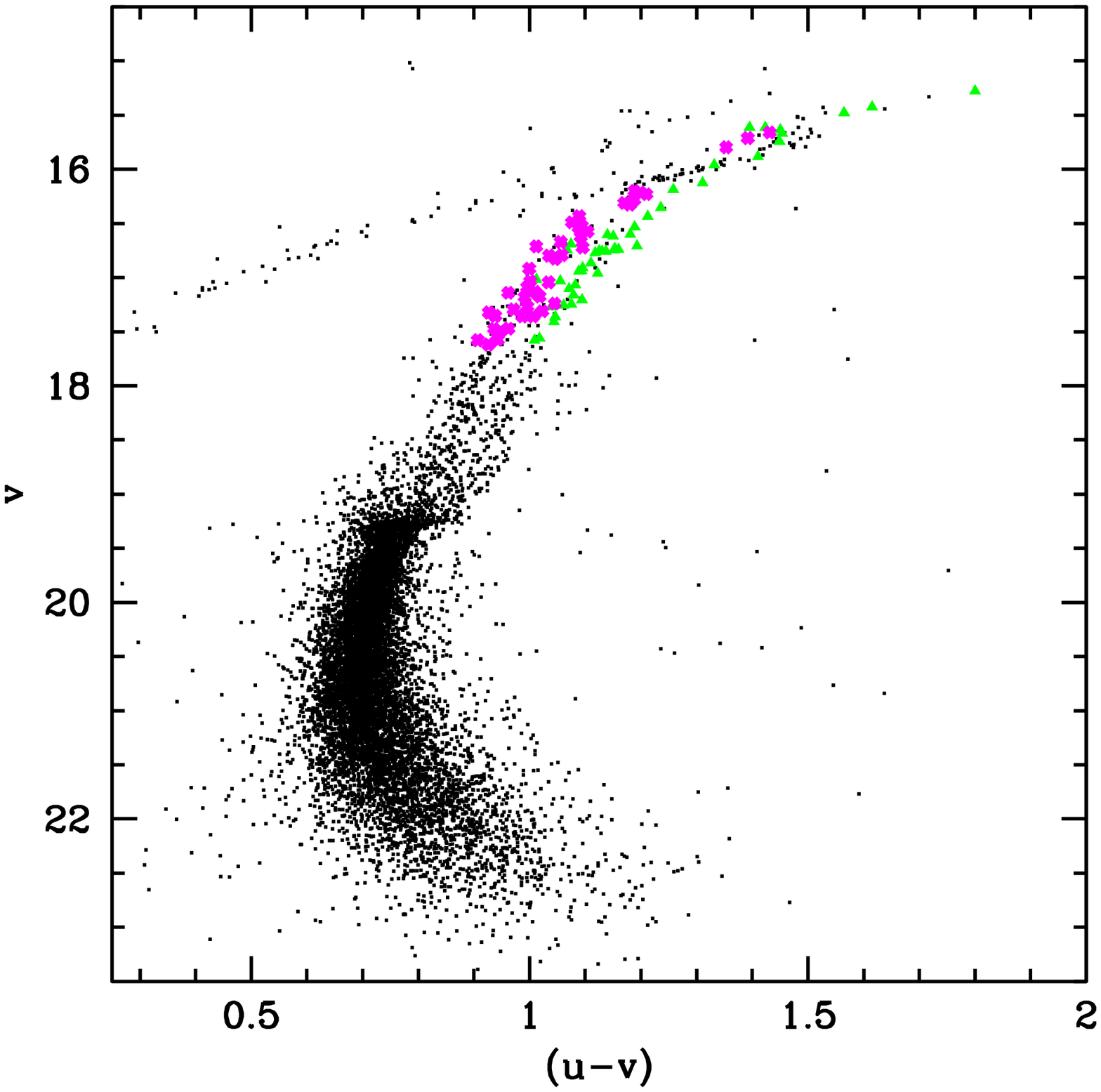}}
\caption{
\footnotesize
CMDs for NGC~288. On the left, the use of our reddest filters shows narrow, well-defined evolutionary sequences, useful to find GC parameters like distance and absolute reddening using isochrone fitting techniques, but it does not allow us to infer the presence of different populations. Fortunately, the use of our bluest filters, especially $u$ (right panel), makes the presence of the multiple populations clearly visible. Magenta crosses and green triangles represent stars from primary and secondary populations as defined spectroscopically by \citet{ca10}. While in our reddest filters, these different populations are mixed, they are clearly correlated with the photometric separation observed using our bluest filters.
}
\label{figngc288}
\end{figure*}

\begin{figure*}[t!]
\resizebox{\hsize}{!}{
\includegraphics[clip=true]{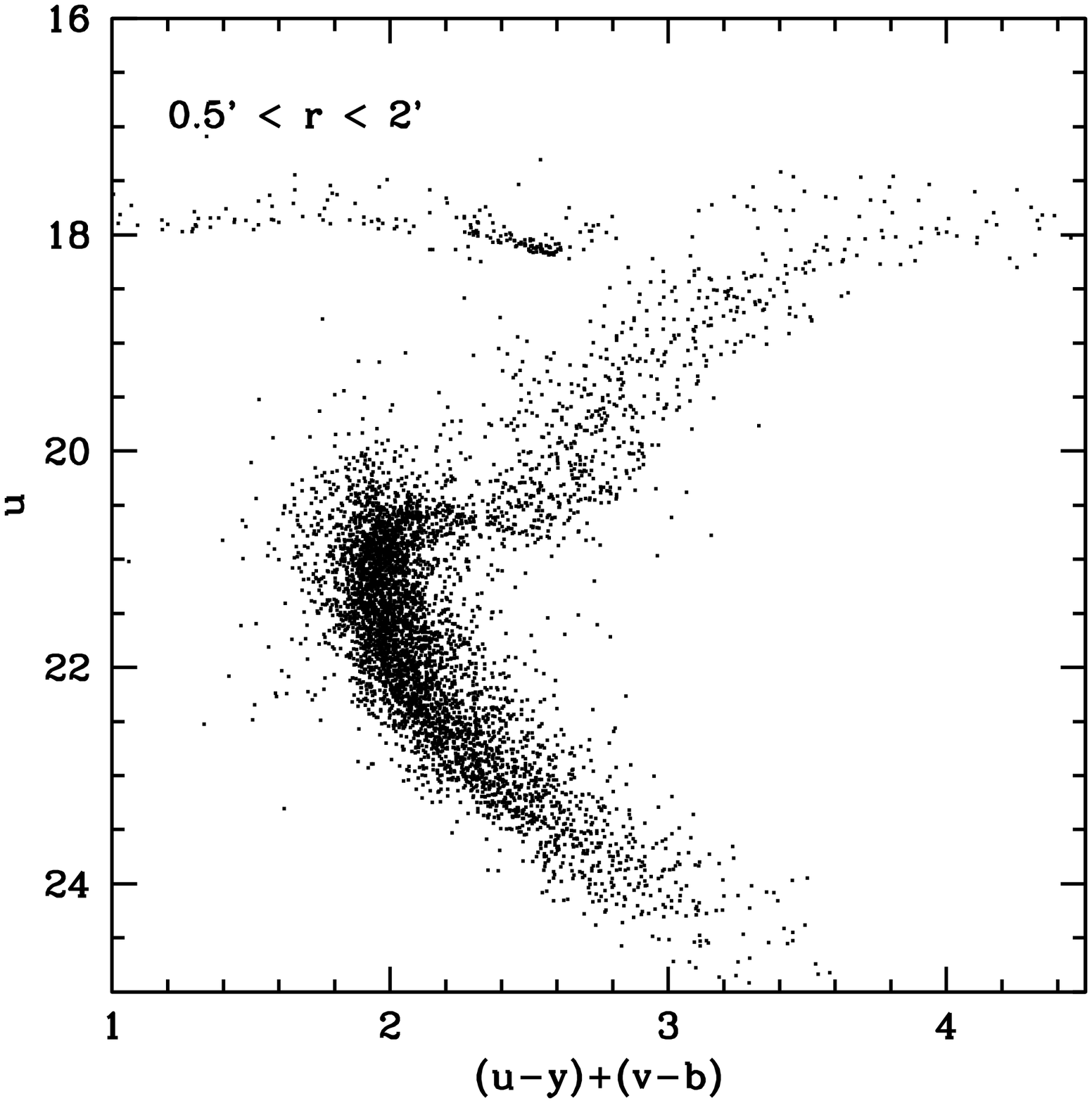}
\includegraphics[clip=true]{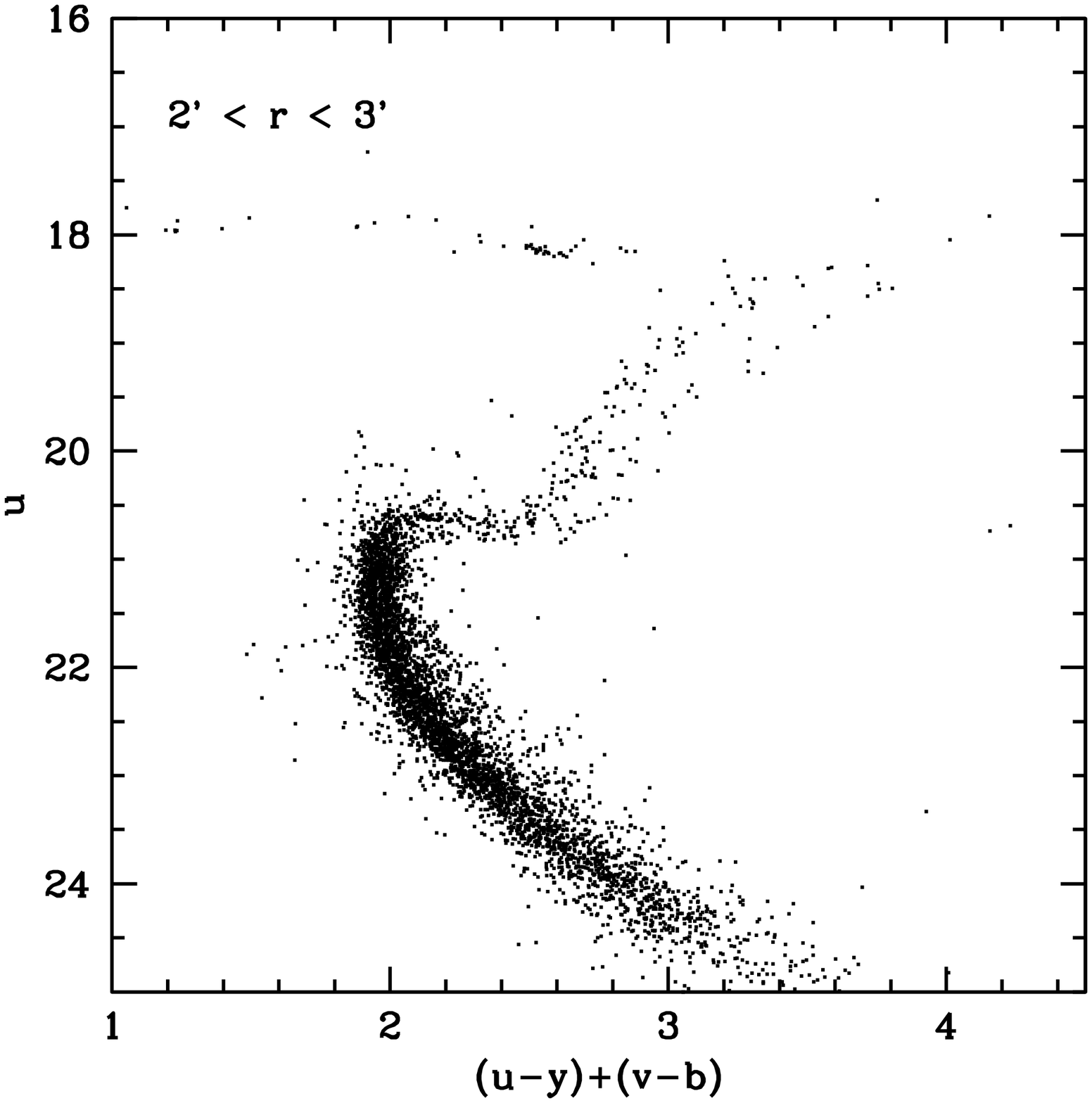}}
\resizebox{\hsize}{!}{
\includegraphics[clip=true]{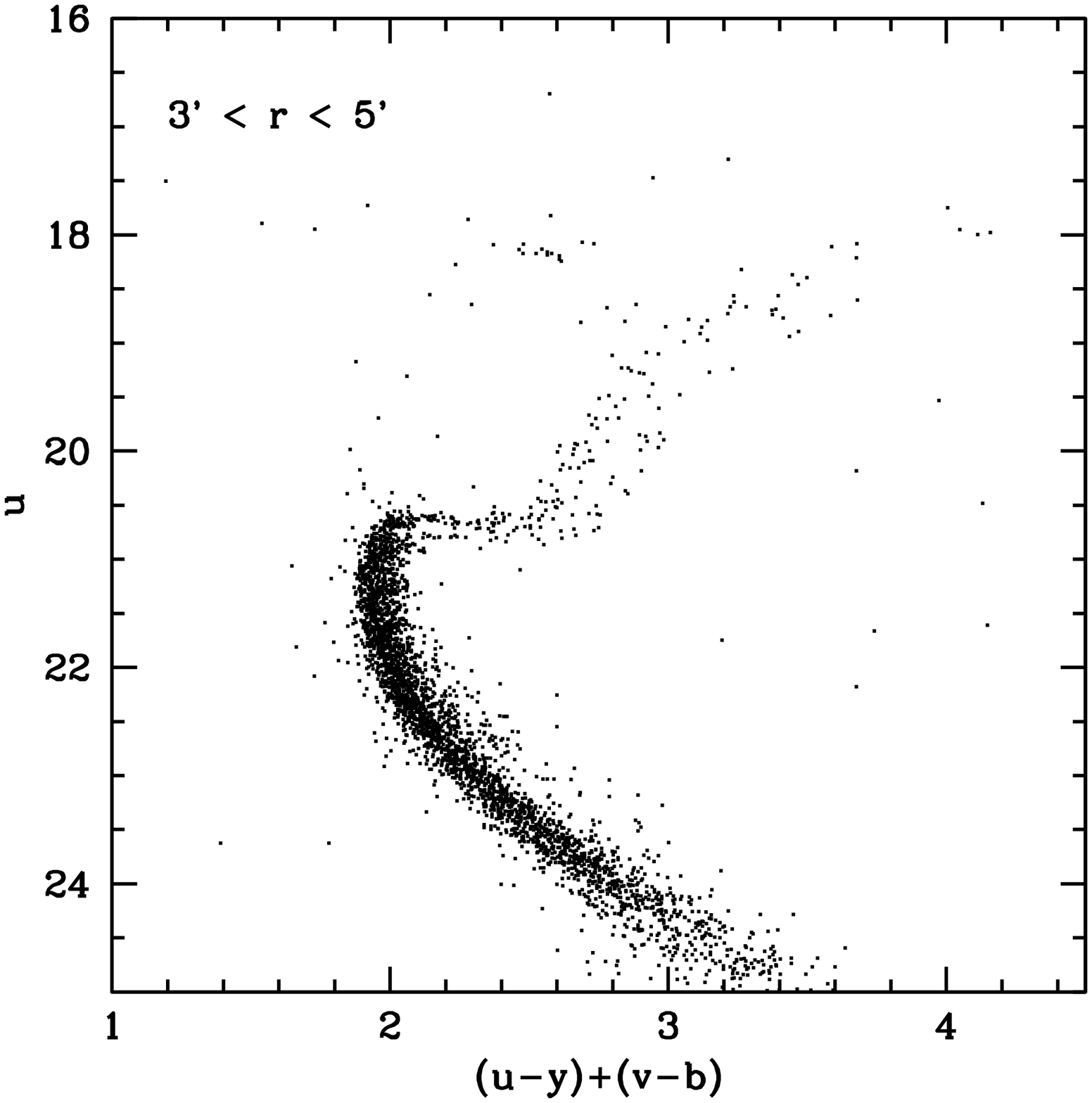}
\includegraphics[clip=true]{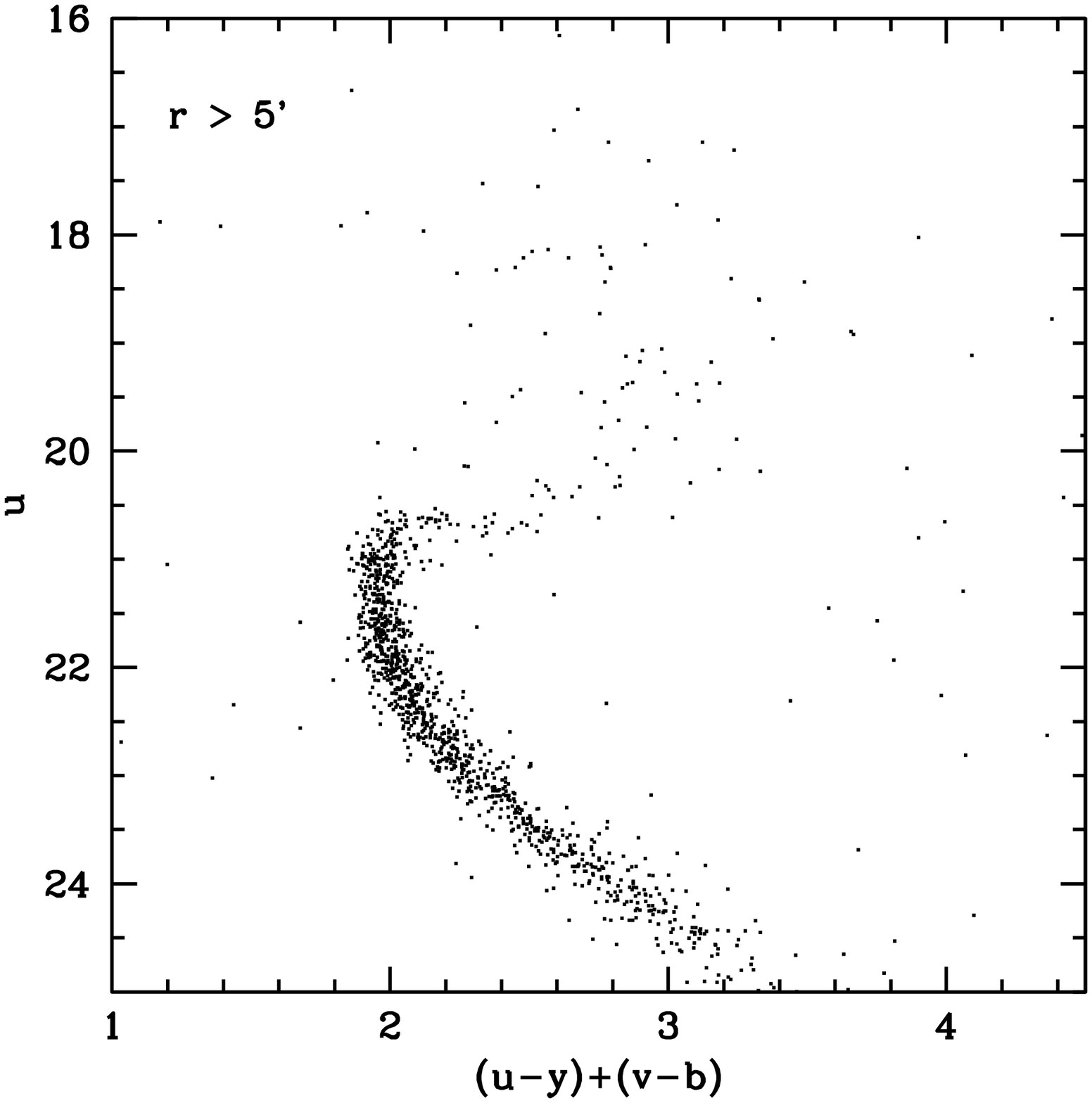}}
\caption{
\footnotesize
CMDs for NGC~1851. The separation in the populations is clearly visible using the color index $(u-y)+(v-b)$ introduced by \citet{la12}, from the SGB up to the RGB. We plot the CMDs for four different radial distances from the center, and can clearly observe the separation in all of them, confirming previous results from \citet{mi09}.
}
\label{figngc1851}
\end{figure*}

We have generated the color-magnitude diagrams (CMDs) for the observed
clusters. The CMDs in our reddest filters (see Figure \ref{figngc288},
left panel) show, as expected, narrow evolutionary sequences, from the
main sequence (MS) up to the tip of the red giant branch (RGB), and do
not provide much information about the presence of multiple
populations, although they can be very useful to extract other types
of information (e.g., differential reddening; \citealt{al12}). But
CMDs with color indices that contain the ultraviolet $u$ passband
(e.g., $(u-v),\ c_1=(u-v)-(v-b),\ \delta_4=(u-v)-(b-y)$) show
broadenings, and even clear separations in some cases (see Figure
\ref{figngc288}, right panel), in their RGBs. These broadenings and
separations are clearly correlated with the population separations
found spectroscopically (see Figure \ref{figngc288}, right panel) in
most of the clusters in our sample for which high-resolution
spectroscopy is available, with only a few exceptions (e.g., NGC~2808,
NGC~7078). A minority of Galactic GCs show variations in Fe and other heavy
elements ~-- e.g., NGC~1851 \citep{yo08}, NGC~6656
\citep{ma09,ab12}~-- and photometrically, they present separations in
their subgiant branches (SGB; \citealt{mi08,pi12}). We also find these
SGB separations in some of the GCs in our sample (see Figure
\ref{figngc1851}), where we can follow them from the cluster center
out to its outskirts.

\section{Summary}

We have recently started a survey to observe a significant sample of
Galactic GCs in the Str\"omgren filter system from the ground, using
the 4.1m SOAR telescope. Our aim is to find and disentangle any
present multiple populations among their stars. First results are very
promising, showing broadenings and separations in all the clusters' RGBs, which
correlate with spectroscopic light element anti-correlations found
for these stars. We have also observed separations in some of the
SGBs, which seem to be present in GCs which have stars with variations
in heavier elements. Our survey is going to let us study these
features from the center of the clusters out to their tidal radii,
allowing us to describe the radial distribution of the multiple
populations in the Galactic GCs with detail.
 
\begin{acknowledgements}
  This project is supported by the Chilean Ministry for the Economy,
  Development, and Tourism's Programa Iniciativa Cient\'ifica Milenio
  through grant P07-021-F, awarded to The Milky Way Millennium
  Nucleus; by Proyecto Fondecyt Postdoctoral 3130552; by Proyecto
  Fondecyt Regular 1110326; by Proyecto Basal CATA PFB-06; and by Anillos
  ACT-86.
\end{acknowledgements}

\bibliographystyle{aa}

\begin{thebibliography}{}

\bibitem[{Alonso-Garc\'ia et al.(2012)}]{al12} Alonso-Garc\'ia, J., et al.\ 2012 \aj, 143, 70

%\bibitem[{Alves-Brito et al.(2012)}]{ab12} Alves-Brito, A., Yong, D., Mel\'endez, J., V\'asquez, S., \& Karakas, A. I.\ 2012 \aap, 540, A3

\bibitem[{Alves-Brito et al.(2012)}]{ab12} Alves-Brito, A., et al.\ 2012 \aap, 540, A3

\bibitem[{Carretta et al.(2009)}]{ca09} Carretta, E.,~et al.\ 2009, \aap, 505, 117

\bibitem[{Carretta et al.(2010)}]{ca10} Carretta, E.,~et al.\ 2010, \aap, 516, A55

\bibitem[{Carretta et al.(2011)}]{ca11} Carretta, E., Bragaglia, A., Gratton, R., D'Orazi, V., \& Lucatello, S.\ 2011, \aap, 535, A121
	
\bibitem[{Cohen \& Kirby(2012)}]{co12} Cohen, J. G., \& Kirby, E. N.\ 2012, \aj, 760, 86

\bibitem[{Gratton, Carretta, \& Bragaglia(2012)}]{gr12} Gratton, R. G., Carretta, E., \& Bragaglia, A. 2012, \aapr, 20, 50

\bibitem[{Grundahl et al.(1999)}]{gr99} Grundahl, F., Catelan, M., Landsman, W. B., Stetson, P. B., \& Andersen, M. I.\ 1999, \apj, 524, 243
  	
%\bibitem[{Grundahl et al.(2002)}]{gr02} Grundahl, F., Nissen, P. E., Briley, M., \& Feltzing, S.\ 2002, in ASP Conf. Ser. 274, Observed HR Diagrams and Stellar
%  Evolution, ed. T. Lejeune \& J. Fernandes (San Francisco: ASP), 228

\bibitem[{Grundahl et al.(2002)}]{gr02} Grundahl, F., Nissen, P. E., Briley, M., \& Feltzing, S.\ 2002, in ASP Conf. Ser. 274, Observed HR Diagrams and Stellar
  Evolution (San Francisco: ASP), 228

\bibitem[{Johnson \& Pilachowski(2012)}]{jo12} Johnson, C. I., \& Pilachowski, C. A.\ 2012, \apj, 754, L38

\bibitem[{Lardo et al.(2012)}]{la12} Lardo, C., et al. 2012, \aap, 541, A141

\bibitem[{Marino et al.(2009)}]{ma09} Marino, A. F., et al. 2009, \aap, 505, 1099
 
%\bibitem[{Marino et al.(2011)}]{ma11} Marino, A. F., et al. 2011, \apj, 730, L16
 
\bibitem[{Milone et al.(2008)}]{mi08} Milone, A. P., et al. 2008, \apj, 673, 241

\bibitem[{Milone et al.(2009)}]{mi09} Milone, A. P., et al. 2009, \aap, 503, 755
	
\bibitem[{Mucciarelli et al.(2012)}]{mu12} Mucciarelli, A., et al. 2012, \mnras, 426, 2889

\bibitem[{Piotto et al.(2007)}]{pi07} Piotto, G., et al. 2007, \apj, 661, L53

\bibitem[{Piotto et al.(2012)}]{pi12} Piotto, G., et al. 2012, \apj, 760, 39

\bibitem[{Sbordone et al.(2011)}]{sb11} Sbordone, L., Salaris, M., Weiss, A., \& Cassisi, S.\ 2011, \aap, 534, 9

\bibitem[{Schechter et al.(1993)}]{sc93} Schechter, P., Mateo, M., \& Saha, A.\ 1993, \pasp, 105, 1342

\bibitem[Skrutskie et al.(2006)]{sk06} Skrutskie, M. F., et al.\ 2006, \aj, 131, 1163

\bibitem[Stetson(2000)]{st00} Stetson, P. B. 2000, \pasp, 112, 925

\bibitem[{Valcarce \& Catelan(2011)}]{va11} Valcarce, A. A. R. \& Catelan, M.\ 2011, \aap, 533, A120

%\bibitem[{Villanova et al.(2010)}]{vi10} Villanova, S., Geisler, D., \& Piotto, G.\ 2010 \apj, 722, L18

\bibitem[{Yong et al.(2008)}]{yo08} Yong, D., Grundahl, F., Johnson, A., \& Asplund, M.\ 2008, \apj, 684, 1159
\end{thebibliography}

\end{document}